# Microdroplets Fail to Retain Exhaled Volatile Biomarkers within a Single Breath

**Author List:** Amio Pronoy Das Ritwik[1,2], Yamin Mansur[1,2], Jingcheng Ma[1,2,3,4,5]*

**Author Affiliations:**

[1]Department of Aerospace and Mechanical Engineering, University of Notre Dame, Notre Dame, Indiana, United States, 46556

[2]Bioengineering Graduate Program, University of Notre Dame, Notre Dame, Indiana, United States, 46556

[3]NDNano, University of Notre Dame, Notre Dame, Indiana, United States, 46556

[4]Berthiaume Institute for Precision Health, University of Notre Dame, Notre Dame, Indiana, United States, 46556

[5]Mike and Josie Harper Cancer Research Institute, University of Notre Dame, Notre Dame, Indiana, United States, 46556

***Corresponding Author:** Jingcheng Ma.

**Contact Email:** jma9@nd.edu

## ABSTRACT

Exhaled breath condensate (EBC) contains volatile metabolites and is promising for non-invasive disease diagnosis, but after decades of research spanning over 100 biomarkers and 10 diseases, no EBC-based test has reached clinical use. The measurement variability that can span orders of magnitude, far exceeding the clinically required 10%, has long been attributed to biological factors. Here, we reveal a fundamentally different origin: the collected microdroplets themselves fail to retain volatile biomarkers. By isolating volatile co-condensation and transient evaporation from biological interference, we show that EBC microdroplets smaller than 100 μm lose clinically significant volatile content within a single breath cycle. This challenges the implicit assumption underlying decades of EBC research, that condensate faithfully reflects airway lining fluid. We develop and validate a physics-based model that predicts this loss across disease-relevant biomarkers and establishes the conditions for reliable EBC sampling. This work reframes EBC variability as a solvable engineering problem rather than an inherent biological limitation.

## MAIN TEXT

## INTRODUCTION

Exhaled Breath Condensate (EBC) is a biofluid collected by condensing exhaled moisture on a cold surface. EBC contains a wide range of volatile biomarkers and has strong potential for non-invasive disease diagnosis and management especially when combined with recently developed biosensors and bioelectronics (1–5). Important biomarkers include inflammatory lipid mediators, oxidative stress markers, cytokines, and metabolites (2, 6, 6–8). Since the first report in 1979 (9), four decades of research have investigated over 100 distinct biomarkers across more than 10 diseases (9), and numerous samplers and biosensors have been developed (10–12). Despite this, no EBC-based test has been approved for clinical use (13, 14).

The major barrier is now widely recognized to be the large measurement variability, which prevents the establishment of reliable diagnostic thresholds (15). Clinically, variability is expected to be within 10-15% CV (16), yet EBC measurements can span orders of magnitude, far exceeding established biofluids such as blood or urine (**Figure 1a, 1b** and **Supporting Table S1**) (17–19, 19–46). This has long been attributed to physiological differences and biochemical contaminations (47–49), while the role of EBC microdroplet phase change behavior including condensation and transient evaporation, as a source of measurement artifacts, remains unstudied (50–54).

EBC sampling involves a complex phase change process encompassing both water and volatile organic compounds (VOCs) (55). Two physical processes are of particular concern. The first is co-condensation: if only water nucleates below its dew point, it remains unclear whether non-miscible biomarkers such as hexane and toluene, are continuously captured, or lost to the gas phase (56–59). The second is the transient evaporation of VOCs from droplets residing on the collector surface or during sample processing (60–64), a concern supported by the fact that pure VOC droplets of approximately 1 mm can fully evaporate within one minute, much shorter than typical EBC collection times (**Figure 1c**).

Comprehensive physical models for EBC co-condensation and evaporation are lacking, due to the complex phase-change physics underlying these phenomena (65, 66). Even for single-phase water, phase change dynamics involve coupled diffusion and convection (67), and are highly sensitive to surface properties (68), droplet size (69), and gas-phase conditions (70). The presence of low-abundance VOCs further complicates this through solutal Marangoni convection (71) and Stefan flow (72). While multicomponent droplet evaporation has been previously studied, existing work largely focuses on the whole-droplet behavior, and does not decouple the behavior of water and VOC transport. Critically, EBC-relevant concentrations fall in an extremely dilute regime (9), often far below 1 %, where water is expected to dominate transport processes, yet trace VOC behavior in this limit remains largely unexplored.

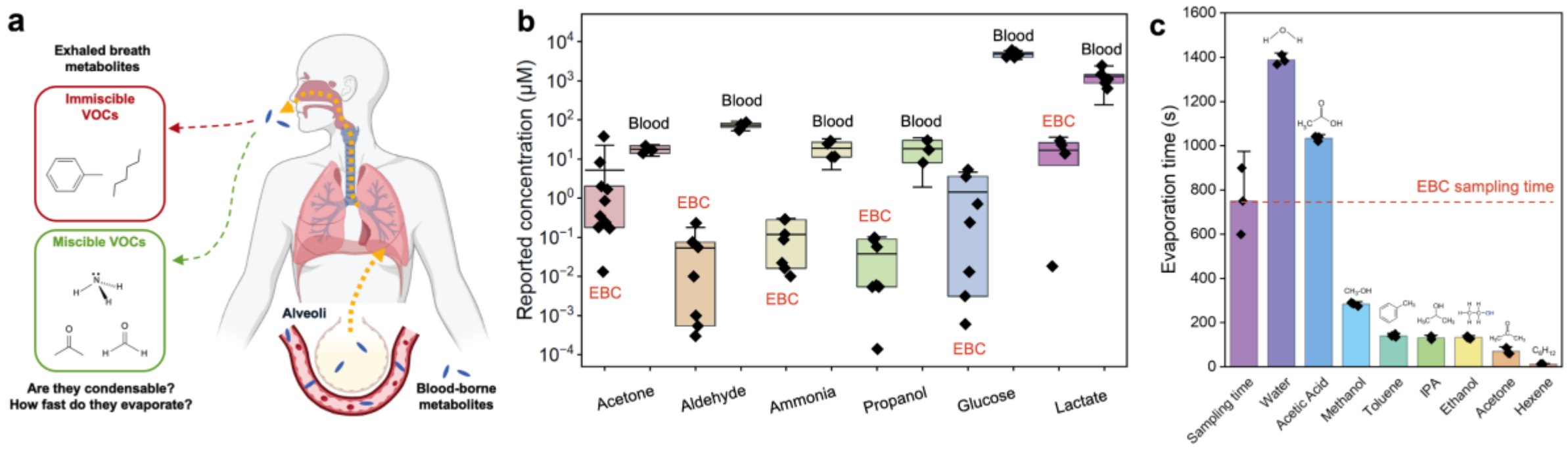

**Figure 1. Motivation and analytical challenges in EBC biomarker measurements. (a)** Schematic of exhaled-breath sampling with miscible and immiscible analytes. **(b)** Literature-reported concentrations for representative biomarkers showing orders-of-magnitude spread between blood (black) and EBC (red), highlighting substantial variability in EBC measurements. The raw data and references are included in **Supporting Table S1**. **(c)** Estimated evaporation times for common VOCs benchmarked against a typical EBC sampling time (red dashed line), emphasizing the potential for evaporation-driven loss during sampling and handling.

In this work, we combine fluorescence imaging, controlled phase-change conditions, and mass spectrometry (73–75) to systematically study VOC losses during two processes: 1) the co-condensation of water and EBC-relevant VOCs, and 2) the evaporation kinetics of dilute mixed droplets (76, 77). Four representative compounds were selected as model systems, including two miscible VOCs (acetone and IPA) and two immiscible VOCs (toluene and hexene). Based on the results of these systems, we developed physical models that can be extended to predict the behavior of other VOCs. The study focuses on the droplet-scale because heat and mass transfer in isolated sub-millimeter droplets are faster and less controlled (78, 79), making them more representative of the conditions contributing to EBC signal loss.

In this study, we show that 1) for condensation, both miscible and immiscible VOCs can continuously accumulate in the outer layer of water droplets at a steady rate, with VOC transport governed by a combination of diffusion and convection; and 2) during evaporation of mixed droplets in the extremely dilute regime (< 5% v/v), VOC evaporation is strongly inhibited compared to pure chemicals. The slow transportation is not limited by liquid-phase transportation, but from gas-phase diffusion. We developed a quantitative physics model that captures these key transport characteristics and used it to systematically predict evaporation losses for up to 12 EBC-relevant compounds as functions of droplet size, orientation, temperature, relative humidity, and VOC physicochemical properties, which is critical for the design of EBC sampling workflow and devices that achieve clinically required measurement consistency.

## RESULTS

**The design and benchmark of fluorescence imaging system for *in-situ* phase change characterization.** Fluorescence imaging is well suited for quantifying VOC behavior in sub-millimeter droplets during both evaporation and condensation, providing resolution in time (1 s), space ($10^{-6}$ m), and the chemical concentration (1% v/v) (80, 81). We developed a customized inverted microscopy system (**Figure 2a**), integrating a Peltier temperature controller and mixture aerosol producer (aerosolization rate: 25 mg/min, pressure: 0.5 bar). A polished sapphire wafer, placed between the cold plate and the condenser wafer, acted as a transparent heat spreader, producing spatially uniform temperature profile at $T = 7$ °C (**Figure 2a** and **Supporting Note S1**).

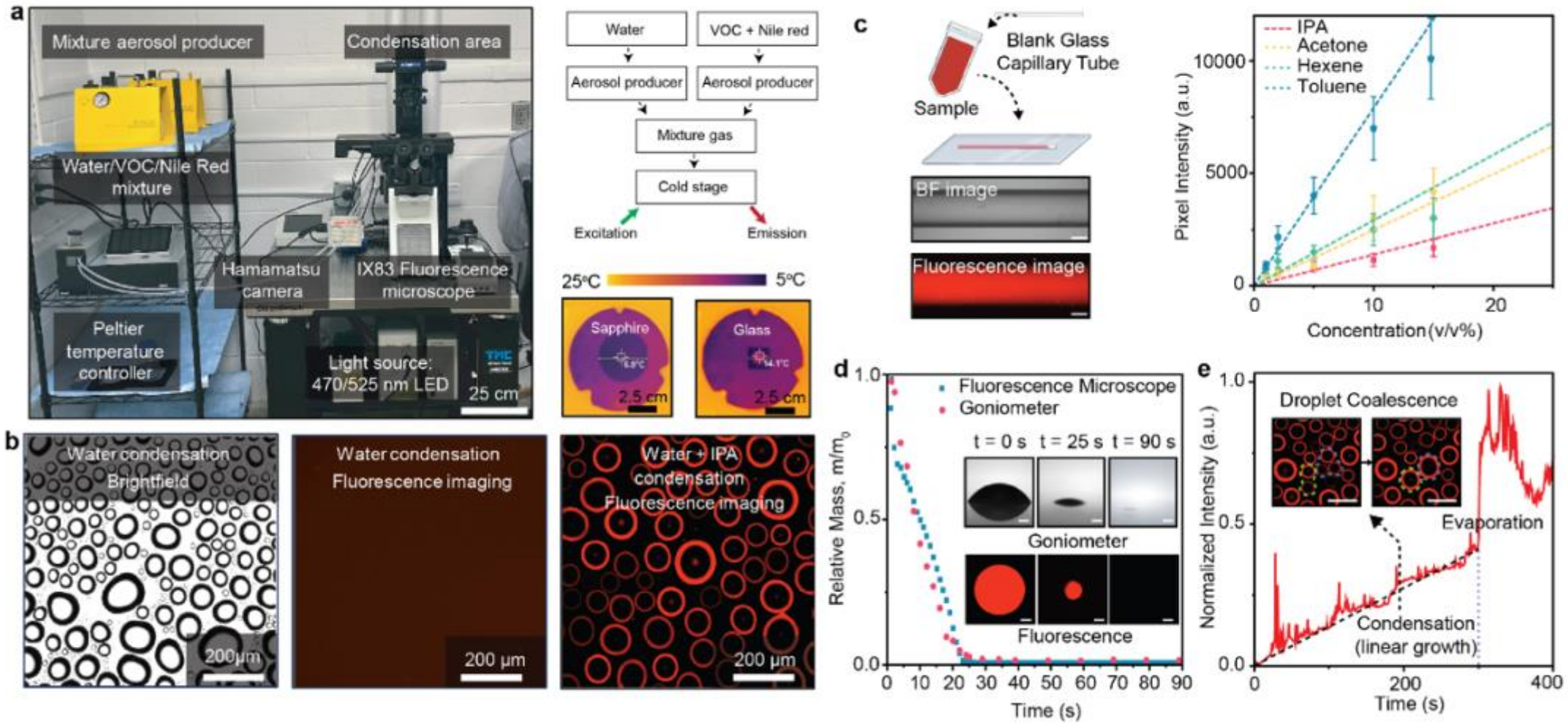

**Figure 2. Visualizing and quantifying VOC co-condensation and evaporation in water droplets using solvatochromic fluorescence microscopy. (a)** Photograph and schematic of the experimental setup: a mixed aerosol/VOC delivery module feeding a temperature-controlled condensation area on sapphire substrates, imaged on an inverted fluorescence microscope to report local polarity via Nile Red response during water/VOC co-condensation. **(b)** Brightfield and fluorescence images of condensed water droplets and water-IPA droplets, demonstrating droplet formation and spatial distribution of VOC-enriched regions during condensation (scale bar shown in panel images). **(c)** Fluorescence calibration curves for Nile Red intensity versus concentration for representative VOCs (IPA, acetone, hexene, toluene) relative to water, and sample preparation workflow (droplet placed under a glass slide and imaged in brightfield and fluorescence) (scale bar: 200 μm). **(d)** Droplet-mass evolution during evaporation measured by fluorescence microscopy and goniometry, confirming consistent time-resolved evaporation kinetics between methods (scale bar: 500 μm). **(e)** Time trace of normalized fluorescence intensity during condensation-driven droplet growth, coalescence and subsequent evaporation (scale bar: 250 μm).

It is essential to decouple the optical signals of trace VOCs from that of the dominant water phase. For this, Nile Red was selected as the fluorescence reporter for its negligible emission in water, enabling selective VOC visualization without interference from the bulk water phase (**Figure 2b**). Fluorescence intensity varies linearly with VOC concentration for all four compounds, allowing optical intensity to be used to quantify VOC abundance and local distribution during EBC sampling process (**Figure 2c** and **Supporting Note S2**). Details of the signal processing methods to remove background fluorescence and optical artifacts are provided in the **Supporting Note S3**. To further validate that fluorescence intensity reflects the VOC content, we independently measured the mass change of a pure acetone droplet using both the top-view fluorescence microscopy and a side-view goniometer (**Figure 2d** and **Supporting Note S4**). The agreement between the two confirms the capability of the platform to capture the VOC concentrations in both single droplets and all the droplets in the field of view. Signal noise in global fluorescence intensity is primarily associated with random droplet coalescence events (**Figure 2e**).

**The thermodynamics of mixture co-condensation.** A key question during EBC sampling is whether VOCs are continuously incorporated into the water condensate at a stable rate. During EBC formation, water is the only species nucleating below its dew point, while VOCs passively participate - raising concern that some VOCs may be effectively non-condensable (**Figure 3a**).

During condensation experiments, we aerosolized pre-mixed water-VOC solution (40 μM Nile Red, 10% v/v) onto a cold stage at 7 °C. A concentration of 10% v/v was chosen to prevent Nile Red clustering, while keeping VOC concentration low enough to avoid exceeding their dew point upon contact with the condensate at the given partial pressure and temperature (**Supporting Note S5**). The condensing glass surface is treated with fluorinated FDTS layer (contact angle: 108°) to ensure reproducible surface energy mimicking hydrophobic Teflon condenser used in many EBC samplers (**Supporting Note 11**, **Supporting Figure S6**).

From **Figure 3b**, we found that: 1) both miscible and immiscible VOCs showed phase separation at low concentration, despite the phase separation only prevalent in immiscible VOCs in bulk solution phase (**Supporting Figure S3**) and; 2) Both miscible and immiscible VOCs grew linearly with water during condensation stage (**Figure 3c**), with VOCs in both cases concentrating at the outer ring. Although we do not quantify the exact fraction of VOCs that enter the liquid phase from gas phase, which was studied in previous work to be about 0.5 (18), these results demonstrated that immiscible VOCs indeed participate in the condensate over time.

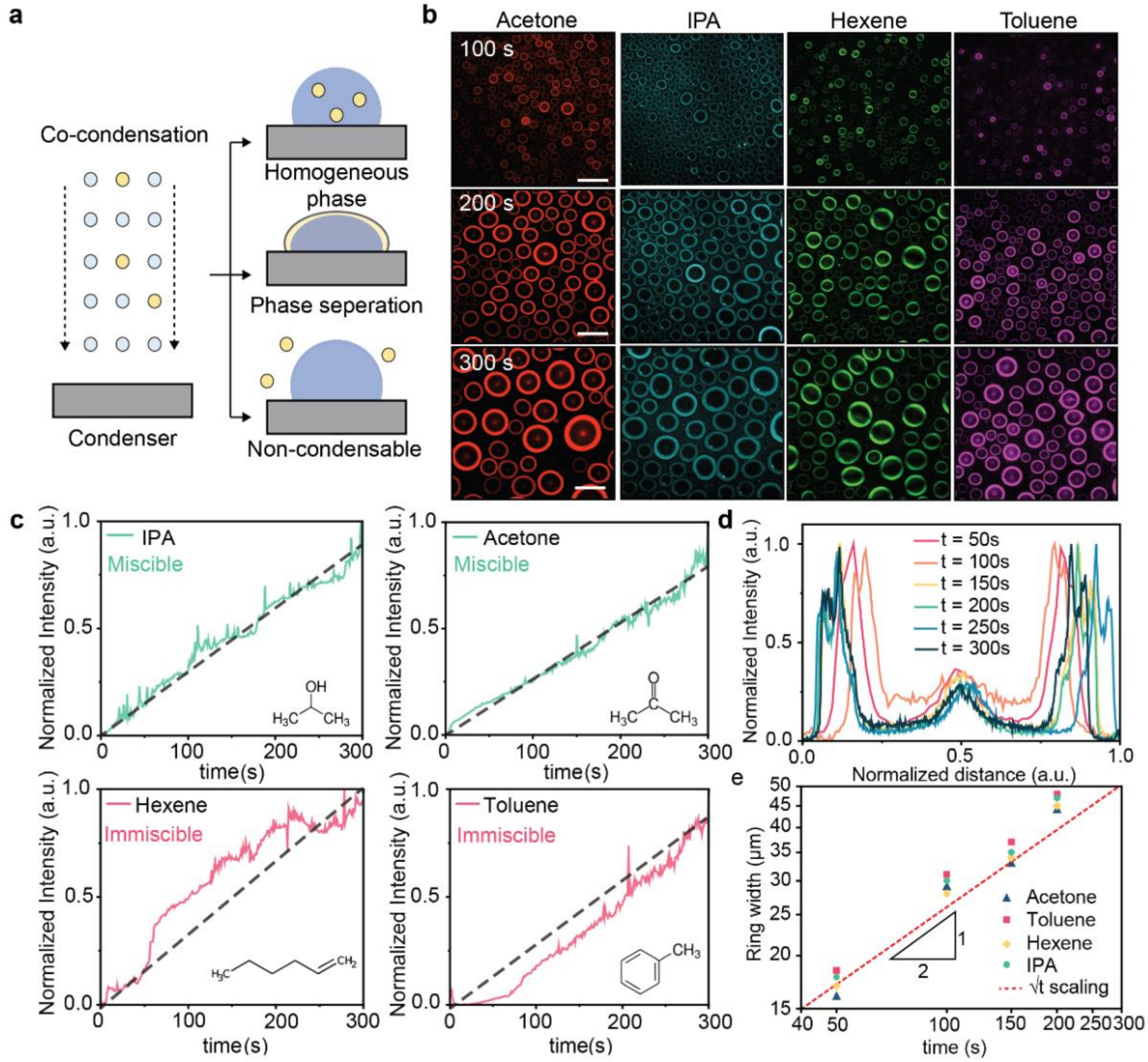

**Figure 3. VOC-dependent co-condensation outcomes revealed by solvatochromic fluorescence imaging. (a)** Schematic of three possible VOC capture outcomes during droplet condensation on the cooled surface: homogeneous incorporation, phase separation into a VOC-rich domain, or negligible capture for effectively non-condensable species. **(b)** Time-lapse fluorescence imaging ($t$ = 0 - 300 s) for representative VOCs (acetone, IPA, hexene, toluene) during condensation, showing similar spatial signatures (scale bar: 500 μm). **(c)** Normalized fluorescence intensity versus time for miscible (IPA, acetone) and immiscible (hexene, toluene) VOCs, highlighting similar accumulation kinetics relative to a linear-growth reference (dashed line). **(d)** Radial intensity profiles at successive timepoints demonstrating the emergence of an annular VOC-enriched region at the droplet perimeter independent of the droplet size. **(e)** Quantification of annular ring width versus time across VOCs, showing a consistent growth trend and supporting the scaling picture for VOC enrichment.

We further extend these observations to assess whether similar behavior is expected for other VOCs. The ring structure was found to be independent of droplet size (**Figure 3d**), and is well described by a physical model in which trace VOCs preferentially adsorb to the water-air interface that lowers surface energy (**Supporting Note S6**). They redistribute primarily by lateral diffusion along interface (**Supporting Note S7**) (82). By treating the interfacial VOC as a diffusing film with an effective surface diffusivity $D_s$, the ring width, $w$, obeys the diffusion-limited scaling $w(t) = 1.9\sqrt{D_s t}$, with $D_s \sim 10^{-12}$ $m^2/s$ consistent with previously reported lateral diffusion coefficient for organic chemicals (83) (**Supporting Note S6** and **S7)**. The model is validated by its ability to describe the evolution of ring width (**Figure 3e**). As the interfacial mechanism is not VOC specific, similar condensation behavior is expected across other EBC-relevant compounds, suggesting that co-condensation is unlikely to account for the reported ~100% measurement variability in EBC.

**The kinetics of transient VOC evaporation from dilute condensates.** If condensation is not the primary source of EBC measurement variability, we hypothesize that transient VOC evaporation from water-dominated condensates is responsible. Two loss pathways are of particular concern during EBC sampling: 1) evaporation from isolated droplets during inter-breath pauses (~ 10 s) (84), where the characteristic droplet size is 1 μm to 1 mm; and 2) evaporation from sample vials during post-processing, over characteristic timescales of 1 s to 1 hour, for millimeter-to-centimeter scale samples. To test this and provide guidance to EBC sampling, a quantitative model that describes VOC evaporation from dilute condensation is required especially at the dilute limit, accounting for the droplet size $R$, droplet contact angle, $\theta$, VOC relative abundance $n_v(t)/n_v(0)$, temperature $T$, relative humidity $RH$, and physiochemical VOC properties including gas diffusion coefficient $D_v$, and saturated pressure $p_v^*$.

Using the fluorescence platform, we studied evaporation of single acetone and IPA droplets (1 μL, room temperature, 30% $RH$) across a range of concentrations, specifically at water-dominating, low concentration limit. We found that at low concentration, although the evaporation rate of the whole droplet remains nearly unchanged, the VOC evaporation rate is strongly reduced (**Figure 4a, 4b**). Both compounds transition from fast to slow evaporation below 5% v/v (**Figure 4c**), well above typical EBC VOC concentrations (ppm-ppb range), confirming that EBC condensates fall firmly in this slow-evaporation regime. However, it remains unclear whether the rate-limiting step originates on the liquid side transport or the gas side evaporation (**Figure 4d**). Immiscible VOCs (hexene, toluene) could not be precisely controlled due to phase separation, but nominal 1% v/v experiments show consistent slow-evaporation behavior (**Figure 4i**).

Liquid-side transport was ruled out due to the apparent convection mass transfer behavior known to provide efficient mass transfer. This can be confirmed by the Marangoni convection-driven ring structure under 3D confocal fluorescence microscope (Leica Stellaris 8 DIVE, **Figure 4e**), and the measured ring structure also follows the predicted dimensionless scaling law from Leveque scaling (**Figure 4f** and **Supporting Note S8**). This result shows that the liquid phase supply to the droplet is not the limiting factor for the observed transition in evaporation kinetics.

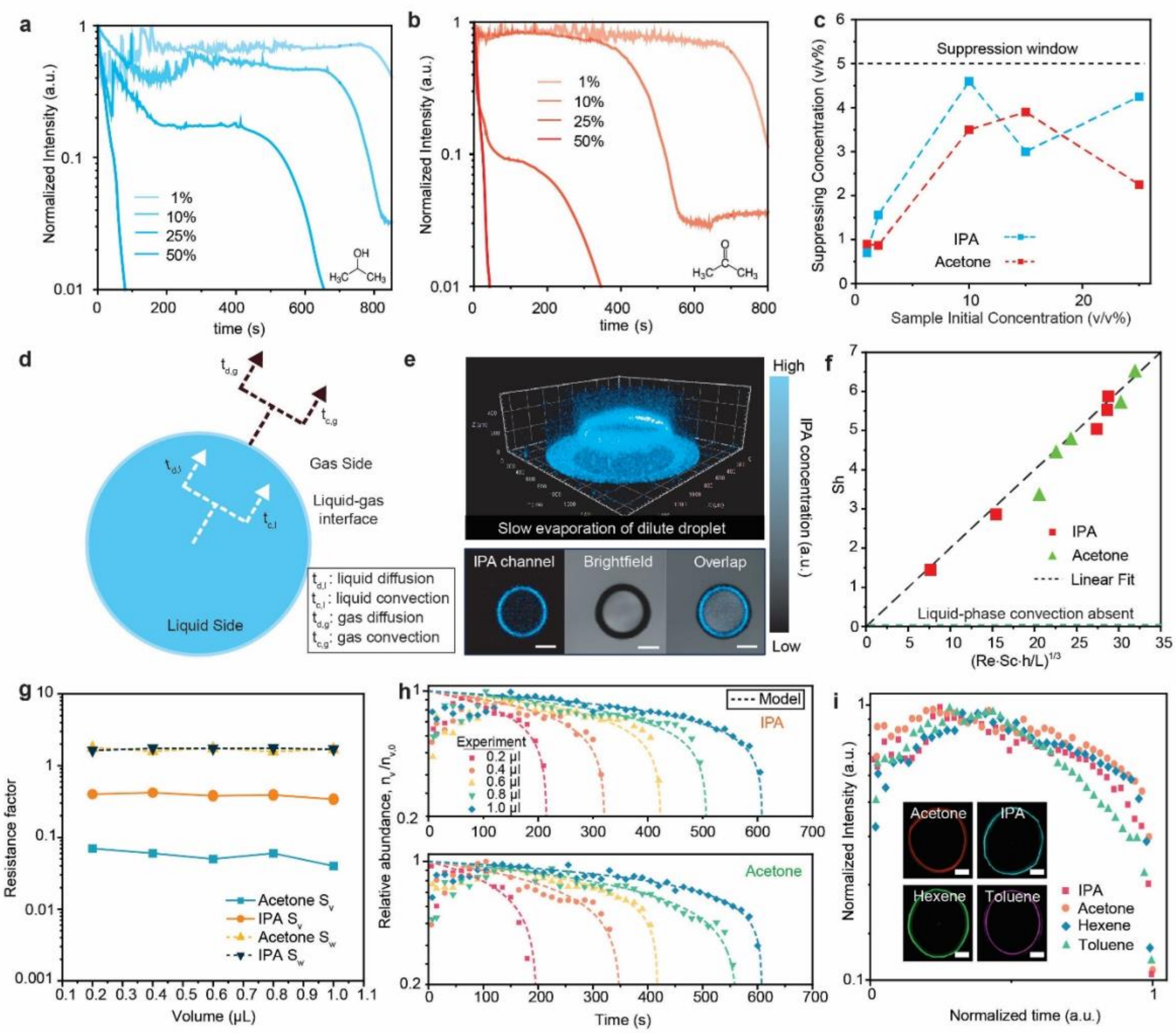

**Figure 4. VOC evaporation from water droplets at low concentration. (a–b)** Time-resolved, normalized fluorescence intensity during evaporation of 1 μL water-VOC droplets (room temperature, 30% *RH*), showing that at higher VOC content IPA (a) and acetone (b) display a rapid initial decay followed by a slower plateau, whereas at 1% v/v only the slow regime is observed. **(c)** The transition into inhibited evaporation is governed by the instantaneous (suppressing) VOC fraction during evaporation rather than the nominal prepared concentration, with the onset consistently occurring below ≈5% v/v for both compounds. **(d)** Schematic of the four candidate VOC transport pathways – liquid diffusion, Marangoni convection, gas diffusion and gas convection – used to identify the rate-limiting resistance. **(e)** Confocal fluorescence characterization of the evaporating droplet ring structure and VOC enrichment, evidencing Marangoni convection within the liquid phase during slow evaporation (scale bar: 500 μm). **(f)** Leveque scaling analysis linking Sherwood number ($Sh$) to ring-scaled Reynolds and Schmidt numbers, confirming that liquid-phase convection is sufficiently active in the droplet. **(g)** Empirical resistance factor $S_{\mathrm{v}}$ and $S_{\mathrm{w}}$, capturing water-mediated gas-phase resistance, supporting a gas-side control of VOC loss in dilute droplets. **(h)** Model-experiment comparison for IPA and acetone across droplet volumes, reproducing trace-regime retention followed by a sharp drop near the drying time. **(i)** Normalized retention curves for four VOCs at matched conditions (1 μL, nominal

1% v/v), demonstrating a unified kinetic signature in the trace regime governed primarily by gas-phase diffusion (scale bar: 250 μm).

We therefore assume that the transition in slow evaporation kinetics is governed by gas-phase transport. A first order gas-side diffusion mass transfer model for both VOC and water in a droplet is developed to provide general description of EBC evaporation. Since water is the dominating phase, we assumed the shape evolution of EBC droplets largely depends on water diffusion and independent of VOC, which is expected to be valid for the low concentration limit. With detailed derivation included in **Supporting Note S9**, the VOC content can be modeled by:

$$n_{\mathrm{v}}(t) = n_{\mathrm{v},0}\left(1 - \frac{tK}{R_0^2}\right)^{\frac{\beta}{K}} \quad \textbf{(1)}$$

Here, $n_{\mathrm{v}}(t)$ is the molar content of VOC at time $t$, and $n_{\mathrm{v},0}$ is the initial molar content of VOC right before evaporation happens at $t$ = 0. The base radius of the droplet is $R_0$. $K$ and $\beta$ are independent parameters that concerns water and VOC physiochemical properties, respectively:

$$K = S_{\mathrm{w}} \times \left(\frac{(0.27\theta^2 + 1.3)}{2f(\theta)}\right)\left(D_{\mathrm{w}}\frac{(1 - RH)p_{\mathrm{w}}^*}{R_{\mathrm{g}}T}\frac{M_{\mathrm{w}}}{\rho_{\mathrm{w}}}\right) \quad \textbf{(2)}$$

$$\beta = \frac{3(0.27\theta^2 + 1.3)S_{\mathrm{v}}D_{\mathrm{v}}p_{\mathrm{v}}^*M_{\mathrm{w}}}{4f(\theta)R_{\mathrm{g}}T\rho_{\mathrm{w}}} \quad \textbf{(3)}$$

Where key parameters include the molar mass of water $M_{\mathrm{w}}$, mass density of water $\rho_{\mathrm{w}}$, saturated pressure of water ($p_{\mathrm{w}}^*$) and VOC ($p_{\mathrm{v}}^*$) at temperature $T$, diffusion coefficient in air of water ($D_{\mathrm{w}}$) and VOC ($D_{\mathrm{v}}$), relative humidity $RH$, and surface wettability that is characterized by water contact angle, $\theta$. Here, $f(\theta) = (2 + \cos\theta)(1 - \cos\theta)^2/4\sin^3\theta$, and $(0.27\theta^2 + 1.3)$ is the Hu-Larson constant (85). The model also includes two dimensionless, empirical gas transfer resistance factors, $S_{\mathrm{w}}$ and $S_{\mathrm{v}}$ for water and VOC respectively. These parameters account for deviations from ideal diffusion-limited evaporation caused by unresolved transport processes near the droplet-air interface. In particular, they represent the effective resistance arising from the thin gas-phase boundary layer above the droplet, where coupled transport of water vapor and trace VOCs occurs. Because this boundary layer contains a dominant water vapor flux and only trace concentrations of VOCs, its detailed structure cannot be directly resolved experimentally. Therefore, $S_{\mathrm{w}}$ and $S_{\mathrm{v}}$ are introduced as correction factors that capture the combined effects of interfacial transport resistance, local concentration gradients, and potential coupling between water and VOC diffusion.

Since gas phase diffusion dominates, we expect $S_w \sim 1$. The only open parameter, $S_v$, is size-independent and experimentally fitted to $S_v < 0.1$ for both IPA and acetone across all droplet sizes (**Figure 4g, 4h**), capturing the low-concentration evaporation physics. This value generalizes across VOC classes, reproducing the behavior of acetone, hexene, IPA, and toluene at matched conditions (1 μL, 1% v/v, **Figure 4i**), confirming the model's predictive capability across EBC-relevant compounds.

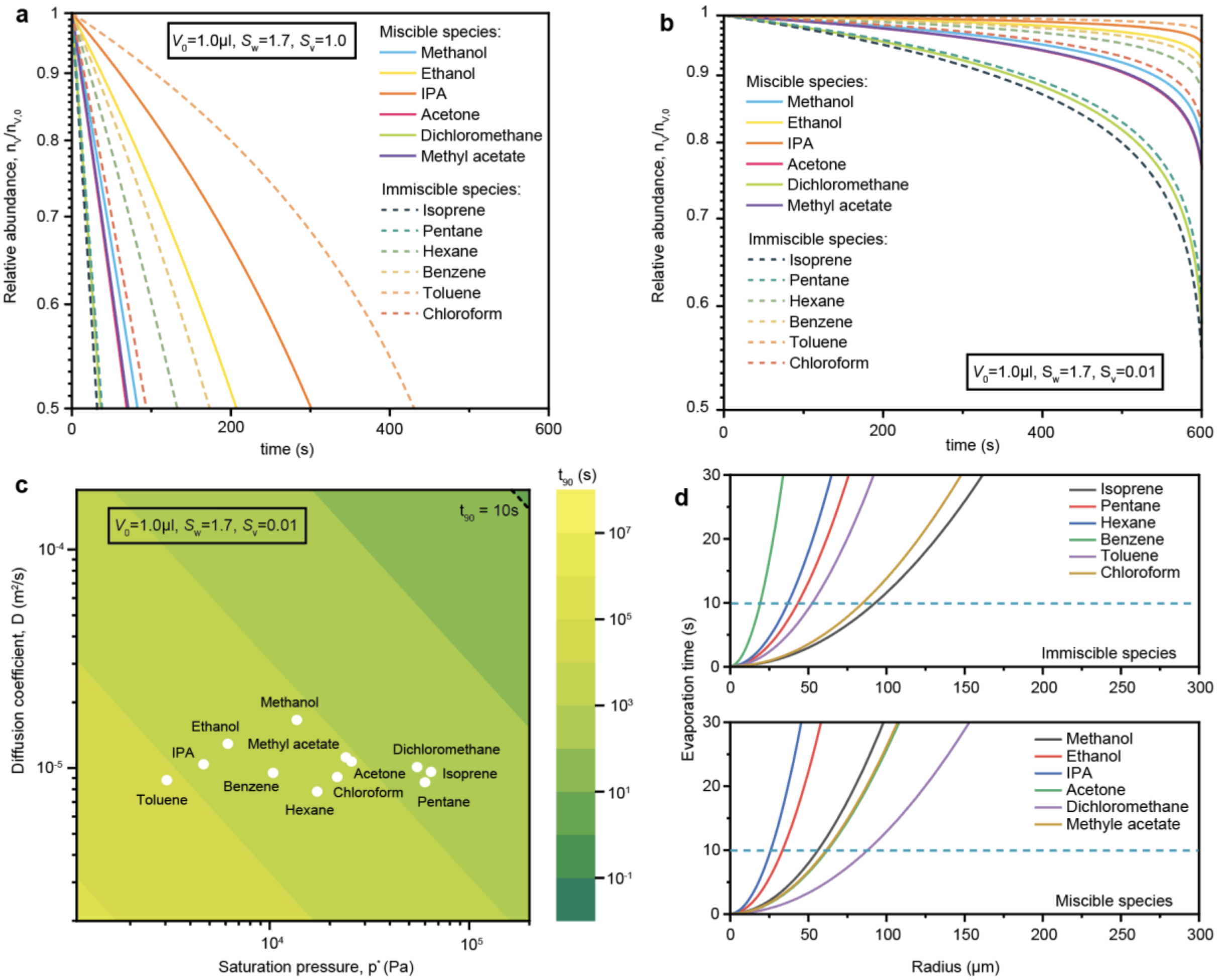

**Figure 5. Predictive model of EBC VOC transient evaporation behaviors in dilute droplets. (a)** Predicted relative VOC abundance during evaporation for miscible and immiscible species under negligible water-phase resistance ($S_v$ = 1.0); evaporation timescales are dominated by each compound's intrinsic volatility and diffusivity. **(b)** Predicted evaporation trajectories under strong water-mediated resistance ($S_v$ = 0.01), consistent with the dilute "trace-regime" observed experimentally. **(c)** Characteristic retention time $t_{90}$ (time for 10% VOC loss) across compounds as a function of saturation vapor pressure and gas-phase diffusivity ($V_0$ = 1 μL, $S_v$ = 0.01); even highly volatile VOCs remain largely retained on minute timescales under dilute, water-dominated conditions. **(d)** Predicted $t_{90}$ versus droplet radius for immiscible (top) and miscible (bottom) VOCs, revealing a compound specific critical radius, $R_{cri}$, below which droplets lose VOCs rapidly within given time window, and contribute minimally to recovered EBC signal. Dashed line: $t_{90}$ = 10 s.

**Theoretical predictions of EBC VOC evaporation as a function of size and time.** Using the predictive model, we analyzed the evaporation behavior of 12 clinically relevant VOCs with different physiochemical properties at high and low concentration. This includes water miscible chemicals (methanol, ethanol, IPA, acetone, dichloromethane, methyl acetate), and immiscible chemicals (isoprene, hexane, benzene, toluene, chloroform, and pentane). At high concentration ($S_\mathrm{v}$ ~ 1), evaporation is fast and has strong dependence on chemical intrinsic properties (**Figure 5a**), consistent with pure chemicals observation (**Figure 1c**). At low concentration limit ($S_\mathrm{v}$ = 0.01), the evaporation behavior converges across all species: slow, diffusion-controlled evaporation over the first 10 minutes, followed by rapid decline as the droplet shrinks (**Figure 5b**), confirming a unified low-concentration evaporation regime.

We define $t_{90}$ as the time for 10% VOC loss, the clinically significant threshold. In the short-time limit, the VOC loss is linear, allowing **Equation 1** to be simplified and express $t_{90}$ as a convenient linear equation for clinical usage:

$$t_{90} \approx R_0^2 \left(\frac{R_\mathrm{g}\rho_\mathrm{w}}{M_\mathrm{w}}\right)\left(\frac{T}{D_\mathrm{v} p_\mathrm{v}^*}\right)\left(\frac{1}{S_\mathrm{v}}\right)\left(\frac{0.14 f(\theta)}{0.27\theta^2 + 1.3}\right) \quad \textbf{(4)}$$

The full derivation of **Equation 4** is in the **Supporting Note S10**.

This expression cleanly decouples the contribution of droplet size, water properties, temperature, VOC properties, the empirical resistance factor, and independence of the relative humidity of the environment. At dilute concentration with $S_\mathrm{v}$ ~ 0.01, even the most volatile compounds show limited evaporation loss for a 1 mm-sized droplet with $t_{90}$ far exceeding 10 seconds (**Figure 5c**). Increasing the size from 1 mm to the standard 1 cm sample vial size will even increase $t_{90}$ by 100-folds, suggesting post-collection handling is not a significant loss pathway.

The dominant loss mechanism is therefore transient evaporation from individual condensate droplets during inter-breath pauses. Since $t_{90}$ depends on $R_0^2$, there must be a critical radius, $R_\mathrm{cri}$, that results in significant sample loss at $t_{90}$ < 10 s. From the theoretical analysis (**Figure 5d)**, we found for all 12 VOCs, $R_\mathrm{cri}$ is on the order of 10 - 100 µm, precisely the size range of dominating droplet populations on flat condensation surfaces (86–89). Droplets below $R_\mathrm{cri}$ can lose a clinically meaningful VOC fraction within a single breath cycle, which we experimentally validate in the following section using acetone as a model biomarker.

**Experimental study of EBC sample loss from transient evaporation.** We experimentally validate two hypotheses derived from our physical model using acetone as model biomarker: 1) the absence of significant long-term sample loss from air exposure for 1-cm sized EBC in vials, and 2) the presence of loss from transient evaporation during condensation (**Figure 6a**). To establish a reference, we conducted human experiments to measure the apparent acetone concentration from human subjects and compared these against artificial EBC (**Supporting Table S7**). A gas-phase acetone concentration of 1 ppm is well-documented from dry gas analysis (90), short-term fluctuations in exhaled breath acetone are limited (91), and the chemical collection efficiency of gas-phase acetone via steam condensation is established at 0.5 (18), giving an expected concentration of 2 mM in a perfect artificial EBC sample with no evaporation loss.

To test hypothesis 1, artificial EBC was prepared at controlled compositions and dilutions (1×, 10×, and 100×) and exposed to ambient air for defined durations ($t_{\mathrm{exp}}$) prior to analysis to mimic open handling condition.(**Figure 6b**). Targeted Gas Chromatography-Mass Spectrometry (GC-MS) was used for quantitative readout of VOC signal across conditions. The acetone peak remained essentially unchanged over all concentration and exposure time, confirming that milliliter-scale EBC is stable during typical open-handling conditions and that extending the exposure interval does not produce a proportional decrease in recovered signal (**Supporting Figures S5 and S6**).

To test hypothesis 2, human EBC was collected and analyzed under identical GC-MS conditions (**Figure 6d**). In contrast to the artificial system, acetone in real EBC appeared near the low end of detectability and was considerably less reproducible, with clear differences observed across three representative samples (**Figure 6e**). **Figure 6f** highlights the droplet-size-threshold retention mechanism that explains both the low signal and variability in human EBC. Rather than forming a uniform film, condensed breath produces a polydisperse droplet population on the cold surface. Droplets below $R_{\mathrm{cri}}$ rapidly lose acetone to the gas phase during inter-breath evaporation and contribute little to recovered signal; only droplets exceeding $R_{\mathrm{cri}}$ retain acetone long enough to be collected. Since breathing conditions vary across individuals and sessions, the fraction of condensate above $R_{\mathrm{cri}}$ shifts accordingly, generating session-to-session variability. This explains why acetone in human EBC is both low in absolute signal and inconsistent across studies, despite being demonstrably stable during post-collection handling.

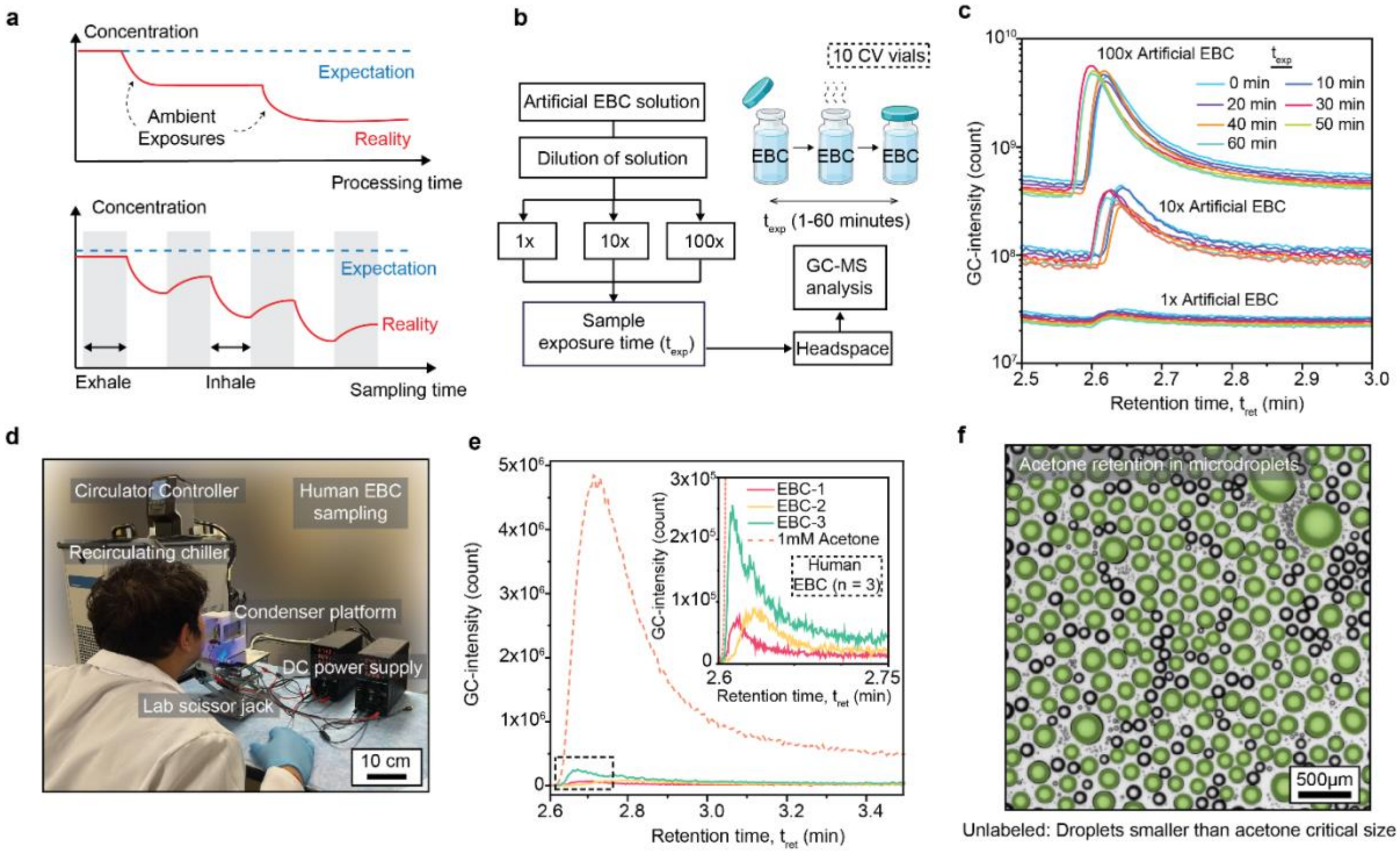

**Figure 6. Validation of EBC stability framework. (a)** Conceptual framework of two hypothesized VOC loss mechanisms: post-collection handling (Hypothesis A) versus loss during collection due to intermittent breathing and evaporation between exhalations (Hypothesis B). **(b)** Artificial EBC validation workflow: defined VOC solution and dilution (1-100×), prescribed open-exposure time $t_{\text{exp}}$ (1-60 min) in capped vials, followed by GC-MS quantification. **(c)** Representative GC-MS chromatograms for acetone in artificial EBC across dilutions and exposure times, showing minimal peak variation and indicating negligible loss during open handling. **(d)** Human EBC collection setup, where exhalation onto a temperature-controlled condenser surface nucleates and captures condensate droplets. **(e)** Targeted GC-MS signal in samples (n = 3), with acetone near the low end of detectability and varies across multiple collection, in contrast to the stability observed in the controlled surrogate matrix. **(f)** Proposed explanation: acetone retention in the polydisperse droplet field depends on droplet size - droplets below a critical radius lose acetone rapidly during evaporation and contribute minimally to recovered signal.

## DISCUSSION

Our study investigates the overlooked physical origins of EBC metabolite inconsistency, including contributions from the VOC-water co-condensation process as well as transient evaporation loss from microscale EBC droplets. Our experiments show that both miscible and immiscible VOCs can be continuously captured during condensation at a steady, linear rate, ruling out condensation failure as a major source of EBC sample loss. Post-collection handling does not contribute to a significant sample loss either: artificial EBC experiments confirm that milliliter-scale samples remain stable over 60 minutes of open-air exposure, because simultaneous water evaporation creates a strong gas-phase resistance that slows VOC loss at dilute concentrations. Our study reveals that transient evaporation of VOCs from microscale condensate droplets between breathing cycles is a physically significant and previously unrecognized contributing mechanism to VOC measurement variability in EBC. A physical model is proposed that predicts a critical droplet radius of 10 - 100 μm, below which droplets lose a clinically meaningful fraction of their VOC content (~10%) within a single inter-breath pause (~10 s). Since this size range dominates the droplet population on typical condenser surfaces, and since droplet size shifts with breathing pattern and individual physiology, the fraction of condensate that actually retains VOCs changes from session to session, which can contribute to the order-of-magnitude spread reported in the literature. We note that EBC measurement variability has multiple recognized biochemical sources beyond the phase-change mechanisms studied here, including salivary contamination, oral cavity contributions, and esophageal gas reflux. Our study does not claim that transient evaporation is the sole or dominant source of variability but rather demonstrates that it is a physically significant mechanism that has not been previously identified or quantified. The quantitative framework developed here provides actionable guidelines for EBC sampler design that can reduce variability below clinically allowed 10% level.

## RESOURCE AVAILABILITY

All data are available to the readers upon reasonable request.

## AUTHOR CONTRIBUTION

The study was conceived by A. P. D. R. and J. M. The experimental platform is designed and built by J. M. and A. P. D. R. Experimental measurements were performed by A. P. D. R. and Y. M. Theoretical modeling was performed by A. P. D. R. and J. M. The manuscript was prepared by A. P. D. R., and J. M., with contributions and final approval from all authors.

## COMPETING INTEREST STATEMENT

The authors declare no competing interests.

## ACKNOWLEDGEMENTS

The 3D confocal imaging for the droplet was carried out in part in the Notre Dame Integrated Imaging Facility, University of Notre Dame using The Zeiss LSM 980 confocal microscope. We thank Dr. Sara Cole for the knowledge and expertise as well as time towards this research. We thank the Materials Characterization Facility (MCF) for the use of the Triple Quad GC-MS. The MCF is supported by Notre Dame Research. We also thank Dr. Ian V. Lightcap from the University of Notre Dame Materials Characterization Facility (MCF) for being very helpful on the assist in GC-MS usage. The work is supported by the University of Notre Dame startup grant and the Technology Development Fund from the Berthiaume Institute for Precision Health.

## DECLARATION OF GENERATIVE AI AND AI-ASSISTED TECHNOLOGIES

Generative AI is used to polish the language of the Manuscript and the Supporting Information during the manuscript preparation stage.